# Compression of colloidal monolayers at liquid interfaces: *in situ* vs. *ex situ* investigation

*Keumkyung Kuk,[a] Vahan Abgarjan,[a] Lukas Gregel,[a] Yichu Zhou,[a] Virginia Carrasco Fadanelli,[b] Ivo Buttinoni,[b] and Matthias Karg [a*]*

[a]Institut für Physikalische Chemie I: Kolloide und Nanooptik, Heinrich-Heine-Universität Düsseldorf, Universitätsstr. 1, 40225 Düsseldorf, Germany

[b]Institut für Experimentelle Physik der kondensierten Materie, Heinrich-Heine-Universität Düsseldorf, Universitätsstr. 1, 40225 Düsseldorf, Germany

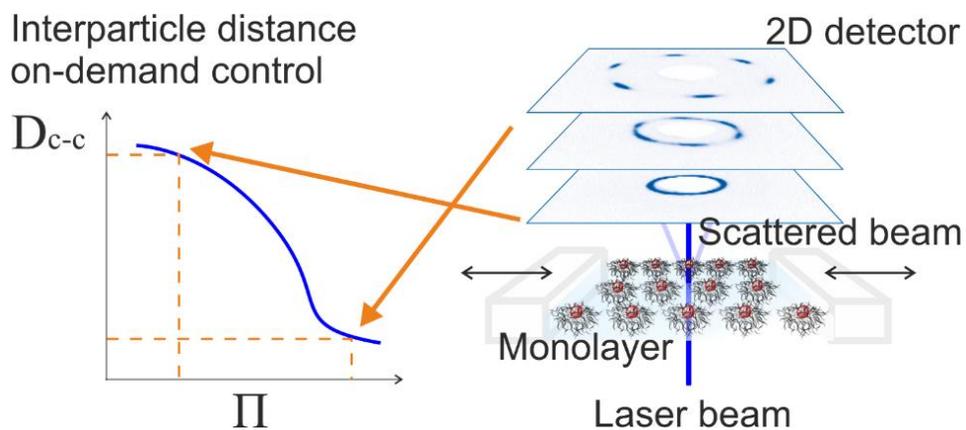

**Abstract:** The assembly of colloidal particles at liquid/liquid or air/liquid interfaces is a versatile procedure to create microstructured monolayers and study their behavior under compression. When combined with soft and deformable particles such as microgels, compression is used to tune not only the interparticle distance but also the underlying microstructure of the monolayer. So far, the great majority of studies on microgel-laden interfaces is conducted *ex situ* after transfer to solid substrates, for example, via Langmuir-Blodgett deposition. This type of analysis relies on the stringent assumption that the microstructure is conserved during transfer and subsequent drying. In this work, we couple a Langmuir trough to a custom-built small-angle light scattering setup to monitor colloidal monolayers *in situ* during compression. By comparing the results with *ex situ* and *in situ* microscopy measurements, we conclude that Langmuir-Blodgett deposition can alter the structural properties of the colloidal monolayers significantly.

**INRODUCTION**

Colloidal monolayers at liquid interfaces, namely, micro- and nanoparticle-laden liquid interfaces, are widely used in fundamental and applied studies. Colloidal particles can self-assemble, in fact, in two-dimensional materials with properties (e.g. photonic or electronic) similar to those of atomic structures. However, unlike atomic counterparts, the colloidal building blocks can be engineered in terms of the chemical composition,[1-4] shape,[4-7] and morphology,[8, 9] in order to tailor the assembly behavior and spatial arrangement. One of the methods for preparing colloidal monolayers is via confining particles at the flat interfacial plane between two immiscible fluids (e.g. an air/water or oil/water interface).[10] This approach offers great advantages not only for studies of gas-liquid-solid phase transitions as the particle concentration (i.e., the number of particles per unit area) can be tuned *in situ* by means of lateral barriers,[11, 12] but also for scalable fabrications for both planar and curved surfaces with areas ranging from cm$^2$ up to m$^2$ scales.[13-15] In the latter approach, the microstructures at the liquid interface are transferred and deposited on solid surfaces (Langmuir-Blodgett deposition) to obtain dried colloidal films, e.g. for coating or photonic applications.[16-20] In contrast to assemblies of rigid spheres, soft colloidal objects like microgels and nanogels[21] can be deformed, for example, under external compression giving access to richer phase diagrams and complex superstructures.[22, 23]

The structural properties of colloid-laden interfaces are typically extracted from microscopy images by detecting the centers of mass of the colloidal units.[24-28] This "particle-tracking" method quickly becomes time-consuming and computationally demanding in the presence of many particles. Even more importantly, it can be only applied above the Abbe diffraction limit. For example, due to the mostly small sizes synthesized to date, assemblies of coreless and core-shell microgels have been mainly characterized *ex situ* – by looking at dried samples with atomic force or electron microscopes – under the assumption that the structure is unaltered during Langmuir-Blodgett deposition.[8, 11, 12, 29-33] Only recently, *in situ* observation of local regions of the particle-laden interface was achieved via atomic force microscopy.[34]

In this study, we propose an *in situ* method – a Langmuir trough combined with small-angle light scattering (LT-SALS) – to characterize colloidal monolayers at the air/water interface. To demonstrate its versatility, colloids of different morphologies were monitored during compression: silica particles (rigid spheres), poly-*N*-isopropylacrylamide (PNIPAM) microgels (soft spheres) and silica-PNIPAM core-shell microgels (hard core-soft shell spheres). The focus of our study, however, lies on the assembly of the core-shell (CS) microgel system. We first present the results from an *ex situ* structural analysis using Langmuir-Blodgett deposition. Then, we compare these results to an *in situ* analysis performed using LT-SALS as well as fluorescence microscopy. Our results indicate that there are severe structural differences between the microstructures of CS microgels at air/water interfaces and after transfer to a solid substrate. These drying effects are in stark contrast with the widely accepted assumption that the interfacial structure is replicated during Langmuir-Blodgett deposition for microgel type building blocks. We discuss analogies and differences with existing works as well as possible reasons for the observed structural changes during drying.

**RESULTS**

**Core-Shell Microgels**

We prepared monolayers of CS microgels at an air/water interface in a Langmuir trough and studied their structure under compression using *ex situ* light microscopy (**Method 1**), *in situ* fluorescence microscopy (**Method 2**) and *in situ* small-angle light scattering (**Method 3**).

CS microgels possess two relevant length scales: the diameter of the incompressible core (here, silica) and the thickness of the soft, deformable shell (here, PNIPAM). In bulk, these length scales simply define the boundaries of the interparticle interactions. When the microgels are confined and spread at the air/water (or oil/water) interface, the situation becomes more complex because the shells laterally deform at the interface leading to changes in shell morphology and shape, and consequently the total diameter, $D_i$ (interfacial diameter), which is larger than the bulk hydrodynamic diameter, $D_h$.[12, 23, 35, 36] Generally, there are three different scenarios for the spatial arrangement of CS microgels at air/water or oil/water interfaces: At very low number of particles per unit area ($n_P/A$), i.e., for near-zero surface pressures, the CS microgels mostly stay apart in an unordered, fluid-like state. In the second regime, as the $n_P/A$ increases, the microgel shells start to touch (shell-shell contact) more frequently. Finally, in the third regime, the shells are squeezed and/or interpenetrated (core-core contact) until the critical point, where the monolayer buckles, breaks and/or is pushed into the subphase (water). Theoretically, if the energy difference between the partially and fully overlapped shells is small enough, energy minimization is achieved by the overlap of shells in some directions at the cost of other neighboring shells,[37-39] leading to a change in the symmetry of the monolayer. In experimental studies, however, such a symmetry change of the microgel monolayer with increasing $n_P/A$ has only been partially observed.[40] In most cases, core-shell structured microgels[12, 29, 30] and coreless microgels[31-33, 41-43] seem to undergo an "isostructural solid-solid phase transition" characterized by microgels in hexagonal arrangement with "shell-shell" contact versus a denser phase, also of hexagonal order, where the microgels are in "core-core" contact.[11, 44] Note that the interparticle distance in "core-core" contact includes the diameter of the core, $D_c$, as well as the compressed microgel shell. The discrepancies between numerical and experimental studies concerning the phase behavior at interfaces have been ascribed to capillary forces and a highly nonlinear mechanical response of the polymer chains (i.e. the PNIPAM corona) under compression[11, 30, 33, 40]

In the following, we report the same "isostructural solid-solid phase transition" only during *ex situ* characterization (**Method 1**) of monolayers of micron-sized CS microgels. Remarkably, this phenomenon is not observed during *in situ* experiments (**Methods 2** and **3**).

**Method 1 (*ex situ* microscopy)**

The *ex situ* microstructural analysis relies on the microscopic investigation of the colloidal monolayer upon transfer from the liquid interface to a solid substrate followed by drying. This leads to dry, substrate-supported colloidal monolayers. When the transfer to the solid substrate is done continuously while the monolayer is compressed in the Langmuir trough, the monolayer position on the substrate can be linked to the corresponding surface pressure at the liquid interface.[11, 44] In this study, the CS microgel system used for the *in situ* and *ex situ* comparison has a core diameter $D_c = 340 \pm 20$ nm and a total hydrodynamic diameter $D_h = 920 \pm 18$ nm (see **Synthesis Section** for more details). **Figure 1A** shows the measured compression isotherm during the Langmuir-Blodgett deposition along with the corresponding microscopy images. Note that we use a linear color coding from light blue to black linked with low to high surface pressure ($\Pi$) throughout this article. From the mid $\Pi$ regime (16.3 mN/m and higher), the images were taken in dark field mode to facilitate image analysis. The monolayer images at lower $\Pi$ were recorded in bright field mode. With increasing compression, i.e., decreasing available area, $A$, the surface pressure increases continuously. In the low $\Pi$ regime, the CS microgels are not homogeneously distributed over the accessible area (see microscopy images) but rather show hexagonal arrangements with shell-shell contacts and some voids among numerous crystalline domains. This indicates the presence of attractive interparticle interactions despite the large interparticle distances, in agreement with previously reported results from *in situ* and *ex situ* analysis of CS microgel monolayers.[8, 12, 44] As $\Pi$ increases, the crystalline domains grow while the voids close. In the high $\Pi$ regime, we observe the formation of CS microgel clusters in "core-core" contact. The critical surface pressure for the start of this "isostructural solid-solid phase transition" is around 16 - 18 mN/m for the presented CS microgel, which can be also identified both in the splitting of the first peak of the radial distribution functions (RDFs, **Figure 1B**) and in the diffraction patterns of the dried monolayers (**Figure 2**). Although the *ex situ* "core-core" distance should lie within the detection limit of our current SALS setup, the microstructures produce diffuse scattering patterns (**Figure 2C** and **2D)**, instead of revealing two distinctive length scales. This is due to the fact that the "isostructural phase transitions" is only locally isostructural, i.e. the monolayers, on mm² scale, do not show a defined symmetry. The transition is more pronounced for higher values of Π, i.e. the diffuse contribution to the scattering patterns increases with increasing Π. For low to medium values of Π, the RDFs are characterized by the first peak corresponding to the interparticle distance, i.e. center-to-center distance from *ex situ* image ($D_{c-c}^{im,ex}$), whereas for high Π above the critical value (e.g. 23.2 mN/m), the peak splits near $D_h$. **Figure 1C** reflects the appearance of these two distinct interparticle distances, as the value of $D_{c-c}^{im,ex}$ approaches $D_h$.

In summary, the *ex situ* analysis reveals that the CS microgel monolayers undergo an "isostructural solid-solid phase transition" upon compression, in agreement with previous studies.[29, 40]

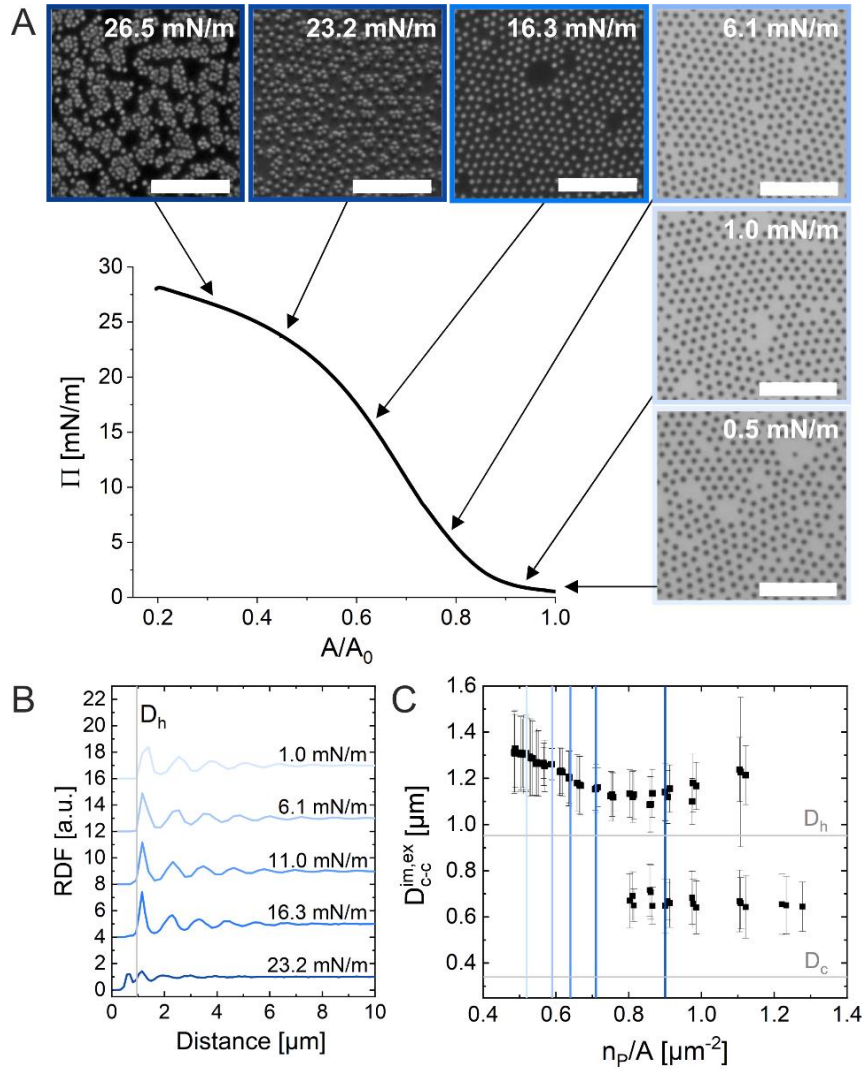

**Figure 1.** A) Compression isotherm of the CS microgels at the air/water interface: Surface pressure (Π) as a function of normalized area (A/A$_0$). The inserted images correspond to optical microscopy images taken *ex situ* using the substrate-supported monolayers obtained from simultaneous Langmuir-Blodgett deposition (dried monolayer). The black arrows indicate the corresponding Π for each microscopy image. The scale bars correspond to 10 μm. B) Radial distribution functions (RDF) for different Π. C) Interparticle distance $D_{c-c}^{im,ex}$ as a function of the number of particles per unit area (n$_p$/A). The colored vertical reference lines indicate the corresponding Π from B).

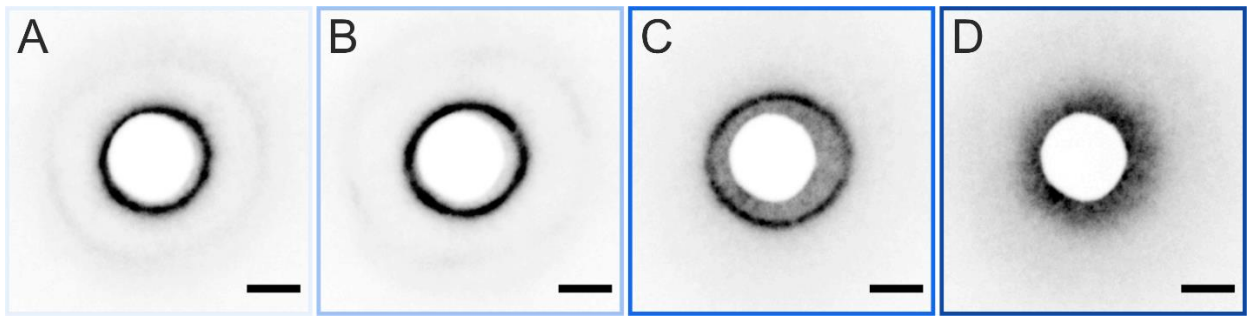

**Figure 2**. Scattering patterns of the dried CS microgel monolayer. The corresponding surface pressures (Π) are A) 0.5 mM/m, B) 6 mM/m, C) 19 mM/m, and D) 24 mM/m. The scale bars correspond to 10 mm.

**Method 2 (*in situ* microscopy)**

*In situ* analysis of the monolayers of CS microgels at the air/water interface under compression was done by combining fluorescence microscopy with a microscopy trough, i.e., a trough equipped with an optical window. **Figure 3A** shows representative microscopy images taken at various values of Π during compression. At near zero Π, we observe clusters of CS microgels due to attractive (capillary) interparticle interactions (see **Figure S1** in **Supporting Information (SI)**), as also reported for other large ($D_h$ > 700 nm) coreless and CS microgels.[8, 44-46] In the regime of low Π, we observe similar microstructures as for the *ex situ* analysis after transfer to a substrate. However, the degree of order appears to be lower at the air/water interface. For medium-to-high values of Π, the comparison with *ex situ* results reveals striking differences: unlike $D_{c-c}^{im,ex}$, the *in situ* interparticle distances, $D_{c-c}^{im,in}$, continuously decrease and the degrees of order increase with increasing Π. An "isostructural solid-solid phase transition" is not observed, in contrast to the assembly behavior reported for other similarly-sized coreless and CS microgels.[29, 44] This becomes even more evident when looking at selected RDFs as presented in **Figure 3B**. The first double peak in the RDFs for higher Π is not present. Furthermore, the higher degree of order is reflected by the large number of distinct peaks in the RDF computed for the highest Π. In contrast to the high Π regime studied in the *ex situ* analysis, the monolayer possesses pronounced long-range order when studied at the air/water interface. **Figure 3C** shows the evolution of $D_{c-c}^{im,in}$ with $n_P/A$ in direct comparison with the values obtained from *ex situ* analysis ($D_{c-c}^{im,ex}$, shadowed area). The data clearly shows a continuous decrease in $D_{c-c}^{im,in}$ with increasing $n_P/A$, indicating a continuous compression of the soft colloidal monolayer. Starting from approximately 1.5 μm, $D_{c-c}^{im,in}$ decreases linearly with increasing $n_P/A$ until a pronounced deviation from the *ex situ* results appears when approaching distances that are close to $D_h$. The final values, at high compression, are slightly larger than half the initial $D_{c-c}^{im,in}$ and lie – until the monolayer buckles – in between the two distinct distances (shell-shell and "core-core") determined by the *ex situ* analysis.

To summarize, the *in situ* measurements using fluorescence microscopy revealed significant differences not only in the microstructure of the monolayer but also in terms of the evolution of the interparticle distance and a noticeable shift in $n_P/A$ for corresponding Π (see **Figure S2A** and **B** in **SI** for more details). For the studied CS microgels, these findings point towards a pronounced drying and/or substrate effect upon transfer by Langmuir-Blodgett deposition, as typically performed for such *ex situ* microstructure analysis. We address this further when discussing the LT-SALS experiments in the next section.

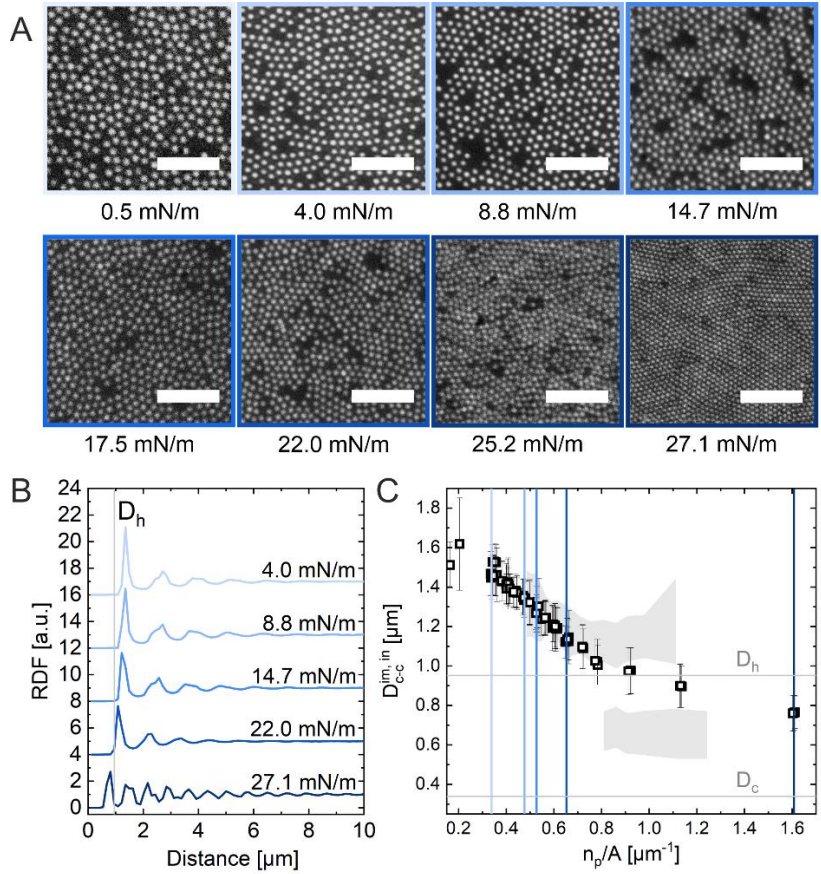

**Figure 3**. A) Fluorescence microscopy images of the CS microgel monolayer at the air/water interface at various surface pressures (Π). The scale bars correspond to 10 μm. B) Radial distribution functions (RDF) for different Π. C) Interparticle distance $D_{c-c}^{im,in}$ as a function of the number of particles per unit area ($n_p/A$). The colored vertical reference lines indicate the corresponding Π from B) and the shadowed area represents the data points from the *ex situ* measurements (**Figure 1C**).

**Method 3 (*in situ* LT-SALS)**

We realized a custom-built setup that combines a Langmuir trough featuring a transparent glass window in the trough bottom (microscopy trough) with a custom-built SALS setup that allows to measure diffraction patterns at high frame rates (up to 30 frames per second, in our case). The details of this setup are provided in the **Experimental Section** and in the **SI**. Furthermore, the **SI** addresses the achievable *q*-range for various laser wavelength highlighting the versatility of the presented LT-SALS method. **Figure 4** shows the compression isotherm along with six selected frames recorded by LT-SALS. The selected frames are correlated to the respective values of Π in the isotherm as indicated by the black arrows. The full video recorded during the compression can be found in the **SI**. The diffraction patterns evolve from a small to a larger ring in a continuous manner in the low to mid Π regime, indicating a continuous decrease in $D_{c\text{-}c}$ in real space. This continuous evolution of the diffraction ring goes on well beyond the critical value of Π where the structural transition was observed in the *ex situ* analysis (Method 1 and **Figure 2**). This is in agreement with the results from *in situ* analysis using fluorescence microscopy (Method 2). As the compression proceeds approaching the high Π regime, the diffraction pattern moves more rapidly away from the center revealing diffraction peaks with six-fold symmetry. After the maximum Π is passed at approximately 34 mN/m, the scattering intensity around the beam stop increases abruptly, indicating the collapse of the monolayer (data are not shown; the collapse can be seen in the Supplementary Video, **SI**).

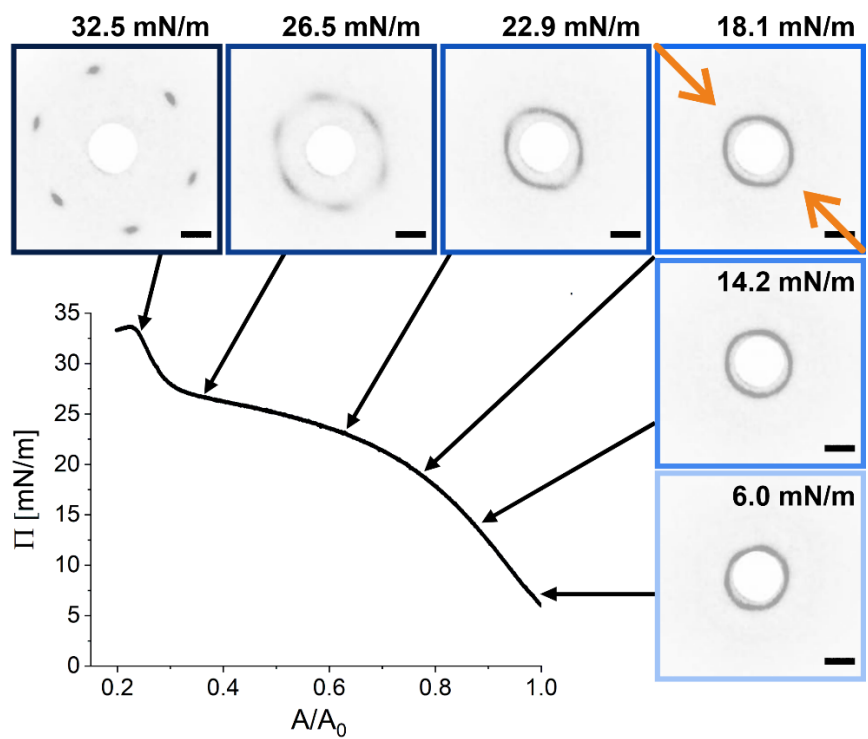

**Figure 4**. Compression isotherm of the monolayer of CS microgels at the air/water interface: Surface pressure ($\Pi$) as a function of the normalized area ($A/A_0$). The images are diffraction patterns of selected frames from a video recorded during the compression (see also **SI**). The black arrows indicate the corresponding $\Pi$ at time of measurement. The greyscale of the images was inverted for better visibility. The orange arrows indicate the compression direction and the white circle at the center is the beam stop. The scale bars correspond to 10 mm (real space dimensions on the detection screen).

**Figure 5A** shows the radial averaging of the intensity of the SALS images as a function of the magnitude of the scattering vector $q$. The resulting scattering profiles show single peaks that correspond to the structure factor of the monolayer. With increasing compression, the position of the structure factor maximum shifts to larger $q$, i.e., smaller real space distances, and its intensity drops significantly in the high $\Pi$ regime. The oval shape of the diffraction pattern in the mid $\Pi$ regime and the increasing full width at half maxima (FWHM) in radial averaging provide interesting insight into the order of the uniaxially compressed monolayer. This, however, is out of the scope of this work and thus will not be discussed here. The calculated real space interparticle distance from the position of the structure factor maximum, $D_{c-c}^{SALS}$ (see **SI** for more details), is plotted against Π in **Figure 5B,** which also contains the values of $D_{c-c}^{im,in}$ from the *in situ* fluorescence microscopy (Method 2) as shadowed area for direct comparison. The two data sets overlap and demonstrate the continuous decrease in interparticle distance with increasing Π. There is no indication of the "isostructural solid-solid phase transition" as observed in the *ex situ* analysis (Method 1). The increase in degree of order during the compression can be monitored by azimuthal averaging at the respective structure factor maximum as shown in **Figure 5C**. With increasing Π, the azimuthal profiles show a transition from a rather isotropic signal to pronounced Bragg peaks at 60° intervals indicating the hexagonal arrangement of the CS microgels in the monolayer. The FWHMs of all six peaks were averaged (FWHM$_{avg}$) and plotted against $D_{c-c}^{SALS}$ in **Figure 5D**. The FWHM$_{avg}$ notably lowers before the $D_{c-c}^{SALS}$ reaches values similar to $D_h$, which supports the previous observation of increasing order.

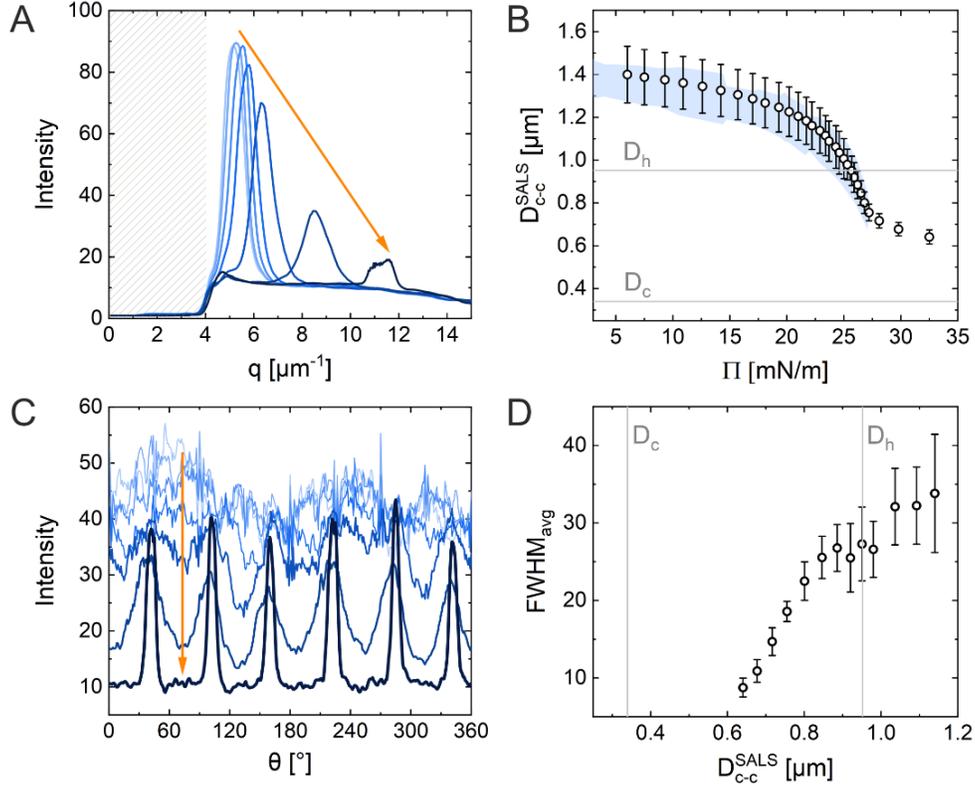

**Figure 5**. A) Normalized integrated scattering intensity as a function of the magnitude of the scattering vector $\vec{q}$ obtained from radial averaging of the diffraction patterns. The shadowed area indicates the area covered by the beam stop. B) Calculated interparticle distances obtained from structure factor analysis of the radially averaged data shown in A), $D_{c-c}^{SALS}$ as a function of surface pressure (Π). The blue-colored area represents the data set from the *in situ* fluorescence microscopy. C) Normalized integrated intensity as a function of the azimuthal angle (θ) as analysis results of azimuthal averaging. D) Width of the Bragg peaks FWHM$_{avg}$ averaged for all six peaks as a function of $D_{c-c}^{SALS}$. The vertical grey lines highlight the core and total CS microgel diameter.

In conclusion, the results obtained using the three Methods (*ex situ* microscopy, *in situ* microscopy and *in situ* LT-SALS) are compared in **Figure 6A**. The grey and blue shadowed areas illustrate the calculated interparticle distances from LT-SALS and *in situ* fluorescence microscopy, respectively, whereas the filled squares correspond to the *ex situ* measurements. The graph highlights the conflict between the *ex situ* and *in situ* analysis of our CS microgel monolayers. The continuous evolution of interparticle distance in monolayers at the air/water interface during the continuous compression was also observed for other CS microgels, as illustrated in **Figure 6B** and **6C,** where the interparticle distance (normalized by the core diameter) is plotted as a function of $\Pi$ for CS microgels with different shell thickness (**Figure 6B**) and overall hydrodynamic diameters ranging from 770 to 1170 nm (**Figure 6C**).

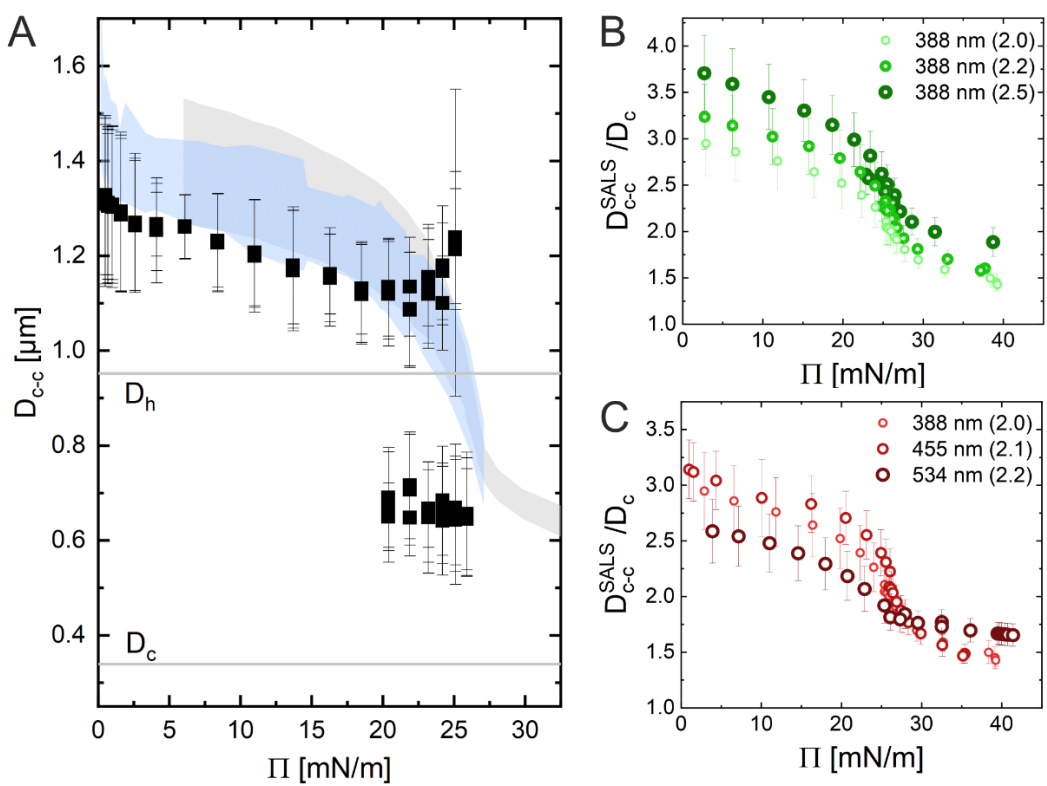

**Figure 6**. A) $D_{c-c}$ versus surface pressure ($\Pi$) plot illustrating the difference between *ex situ* (black squares) and *in situ* (blue and grey shadows) measurements. B) LT-SALS analysis results from core-shell microgels with various shell thickness and C) different overall sizes. The numbers

outside of the parentheses in the legend denote the diameter of the silica core (i.e. 388 nm) and the number inside of the parentheses their shell-to-core size ratio (i.e. 2.0).

**Application of LT-SALS to Other Colloidal Systems**

In this section, we would like to briefly emphasize the versatility of LT-SALS by showing data for two additional representative colloidal systems, i.e., silica particles (as an example of rigid spheres) and PNIPAM microgels without rigid cores (as an example of soft spheres). In **Figure 7**, the compression isotherms are shown along with the diffraction patterns for selected values of Π.

The incompressible nature of the silica particles (diameter measured by transmission electron microscopy, $D_{TEM}$ = 695 ± 22 nm) with a nearly *hard-spheres* interaction potential is well depicted by the steeply increasing Π over relatively small changes in accessible area $A$ (**Figure 7A**). The diffraction images corresponding to different values of Π reveal diffraction rings indicative of polycrystalline microstructures without any preferred domain orientation, as reported for other hard sphere and hard sphere-like systems.[47-49] As such, there are no distinguishable Bragg peaks. During compression, the position of the diffraction ring remains nearly unchanged ($D_{c-c}^{SALS}$ = 719 ± 15 nm). Therefore, the system is characterized by only one length scale from the beginning to the end of the compression as expected for hard spheres in contact. This characteristic length scale, i.e. *in situ* interparticle spacing, of the silica particle monolayer is also present in the *ex situ* microscopic images and the diffraction patterns of the dried silica monolayer as shown in **Figure S3** (**SI**).

The diffraction pattern obtained from the monolayers of the PNIPAM microgels ($D_h$ = 858 ± 41 nm) goes through a transition from a diffraction ring (unordered state) to six distinct Bragg peaks (hexagonally ordered state) near the maximum Π. The monolayers showed rather small changes in Π per area reduced over the course of compression. The experiment was conducted in a highly compressed state not only because the high Π regime is where the structural change is most visible but also because the interparticle distance in the low Π regime is far too large to be resolved by diffraction analysis with our current setup. **Figure 7B** illustrates LT-SALS measurement from the final stage of the compression. The position of the diffraction patterns changes from the edge of the beam stop ($D_{c-c}^{SALS} \approx$ 1690 nm) and moves away from the center to the furthest peak position, i.e., the smallest possible interparticle distance ($D_{c-c}^{SALS}$ = 762 ± 128 nm), although there is only a slight change in Π. The *in situ* fluorescence microscopy (Method 2) at lower Π (0-29.2 mN/m) in **Figure S4** (**SI**) confirms that the interparticle distance at the air/water interface evolves in a continuous manner throughout the compression also for these coreless microgels. Additionally, the contrast in the spatial arrangements between the *in situ* monolayer and the dried microgels (**Figure S5)**, which resembles reported dried monolayer of similarly sized CS microgels,[40] further supports our conclusion that the drying process accompanies structural changes.

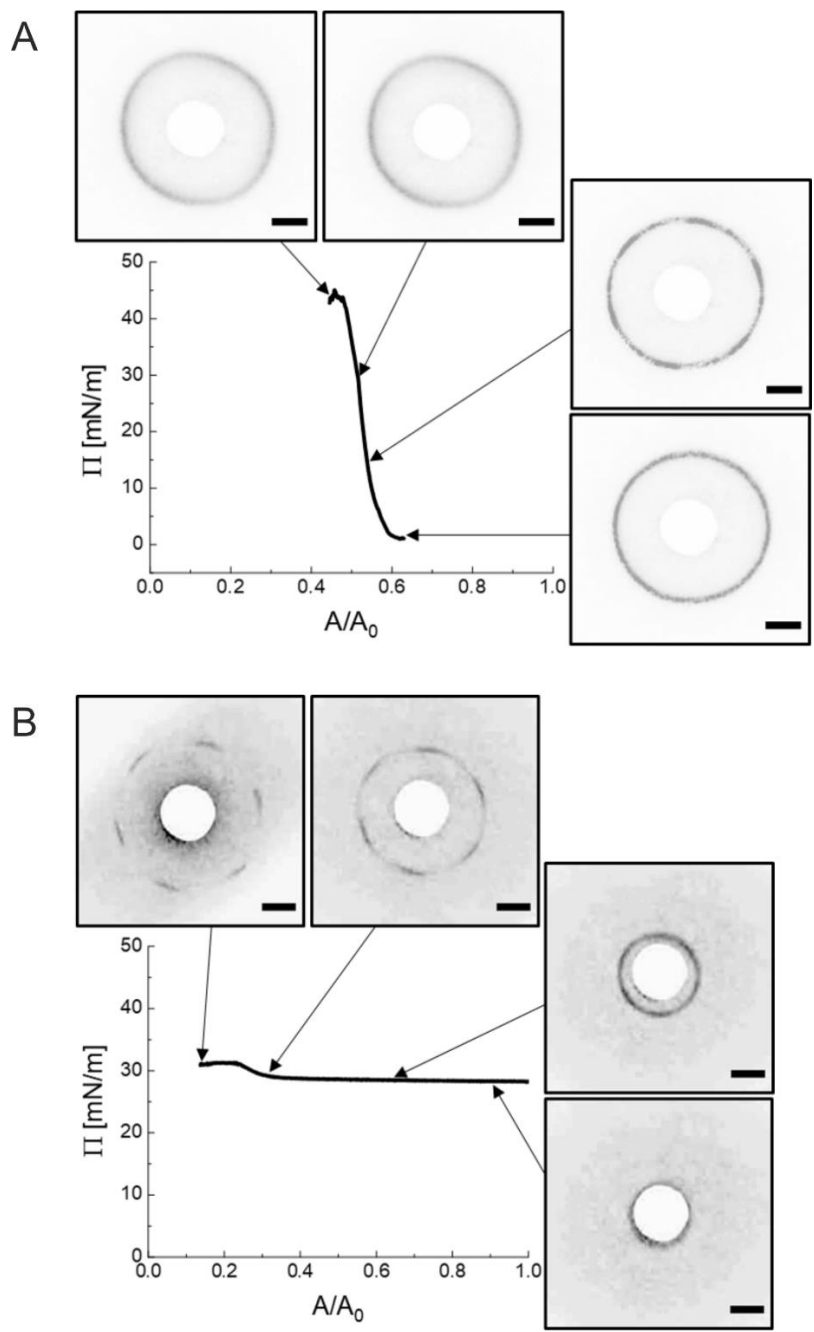

**Figure 7**. A) Π against $A/A_0$, compression isotherm for silica particles along with the diffraction patterns at corresponding Π. B) Π against $A/A_0$, compression isotherm for PNIPAM microgels along with the diffraction patterns at corresponding Π. The contrast of the images was adjusted for better visibility. The scale bars correspond to 10 mm.

**DISCUSSION**

"Isostructural phase transitions" in microgel-laden monolayers under compression have been ascribed to the combination of attractive capillary forces and local failures of the polymer shells, which would otherwise prevent "core-core" contact. Both contributions depend on various parameters including the degree of deformability of the shells (which is mainly determined by the crosslinker density[50]), the size of the core and the shell, the materials, and the overall synthesis protocol. *Ex situ* measurements suggest that microgels with low crosslinker density show a more continuous evolution of the interparticle distance, whereas higher crosslinker densities give rise to "isostructural phase transitions".[29, 31-33] However, we want to note that small microgels with low crosslinker density tend to self-assemble into less ordered structures.[51] Phase transitions seem also to be more likely when large microgels (e.g., $D_h$ = 1450 nm[44]) are used or when the polymer shells are thicker in the case of CS microgels.[29] Nonetheless, there exist still several controversial results; for example, Vogel et al.[52] and Rauh et al.[12] studied CS microgels of similar size, but "isostructural phase transitions" were only observed in reference.[12] Importantly, all these results are based on *ex situ* measurements and only recent works started to provide *in situ* data.[34, 53] In particular, acquiring structural information on statistically relevant areas remains challenging.

In this manuscript, we used *in situ* methods (Method 2 and Method 3) to investigate monolayers of CS microgels of size and crosslinker density similar to references[29, 33, 40, 46] and, to a smaller extent, monolayers made of coreless microgels similar to references.[44, 45] In all cases, our results strongly point towards a continuous evolution of the interparticle distance, i.e. no "isostructural phase transition". The direct comparison with *ex situ* measurements (Method 1) suggests that the "structural transitions" are an artifact of the transfer and/or drying process. As such, the conflicting literature can be partly explained by taking into account the further complexity introduced by the *ex situ* measurement protocol. For example, in contrast to the often applied synchronized Langmuir-Blodgett deposition during compression also depositions at fixed surface pressures were performed.[31] What happens when a microgel-laden monolayer is transferred onto a solid substrate? **Figure 8A** illustrates a sketch of a typical CS microgel, while **Figures 8B1** and **8C1** depict CS microgels with shells of different deformability at the air/water interface. As water evaporates, the microgels approach the substrate and the bottom part of the microgels will start to touch the substrate, most likely causing further deformations as illustrated in **Figures 8B2** and **8C2**. The contact area between the microgels and the substrate and the resulting adhesion depends on (1) the properties of the microgels (e.g., their morphology), the ones of the underlying surface (e.g., its wettability) and the transfer protocol (e.g., deposition speed).[54, 55] As the level of subphase lowers further, the microgels protrude more and more from the liquid film as shown in **Figures 8B3** and 8**C3**, leading to a deformation of the meniscus and attractive immersion capillary forces.[56-58] Although these forces have not been measured experimentally for CS microgels, they qualitatively explain the formation of clusters as monolayers are transferred to the substrate.

We briefly verified that the transfer protocol affects the *ex situ* assemblies by drying CS microgel monolayers with overall hydrodynamic diameters ranging from approximately 500 to 1000 nm[59] (5 mol.% crosslinker density) with two different drying conditions; 'slow' drying at ambient conditions against open air, and 'fast' drying using a heat gun. **Figure 9** shows that structural changes − consistent with an "isostructural phase transition" at the interface − appear only after slow evaporation (blue panels). This observation implies that the microgels have enough time to rearrange when the monolayer is dried slowly under ambient conditions. This is in line with the experimental and theoretical findings of Volk et al.[60] It is noteworthy, however, that the "freezing" of monolayers by fast drying has its limits and will depend on the core dimension, the shell-to-core size ratio, and the crosslinker density. A similar conclusion was drawn by Vasudevan et al.[46] Furthermore, as the temperature influences the microgel fraction in the water subphase and mostly along the vertical direction,[53] differences in the adhesion and contact area with the substrate during drying are expected. The AFM images of **Figure S6 (SI)** reveal such a structural difference between slow and fast-dried monolayers. In particular, the phase images show the difference in contact area on the substrate. Similarly, Bochenek et al. have also reported that the drying conditions have a direct influence on the resulting microstructure.[41, 42]

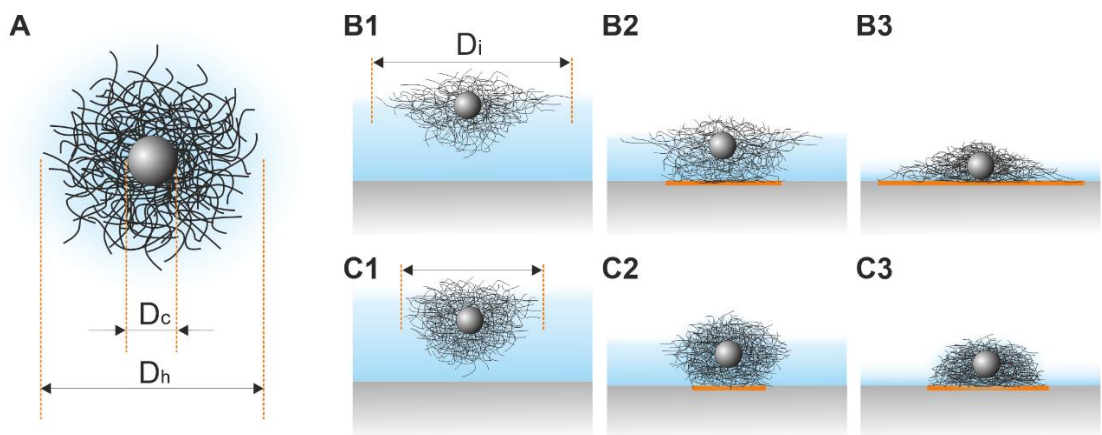

**Figure 8**. A) Schematic illustration of the structure of a CS microgel. $D_c$ denotes the diameter of the core and $D_h$ the hydrodynamic diameter. B1)-B3) CS microgels with high deformability adsorbed at the air/water interface at three different drying stages. $D_i$ denotes the diameter at the interface and the orange area depicts the contact area between the microgel and the substrate. C1)-C3) the same set of illustrations for CS microgels with low deformability.

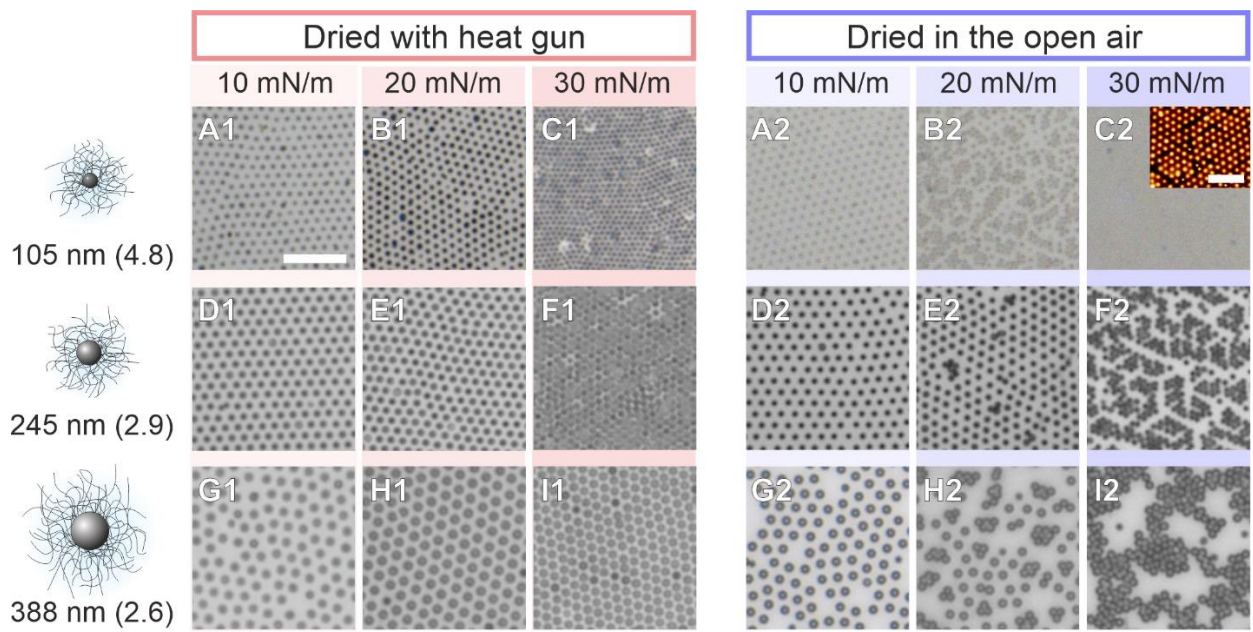

**Figure 9**. Influence of drying conditions of CS microgel monolayers transferred to solid substrates at different surface pressures of 10 mN/m (A), 20 mN/m (B), and 30 mN/m (C). Left panel: Fast drying using a heat gun. Right panel: Slow drying against air at ambient conditions. Shown are results from CS microgels with different core sizes (105, 245, and 388 nm, from top to bottom) and shell-to-core size ratios. The scale bars correspond to 5 μm. The inset in C2 is an AFM image of the corresponding monolayer. The scale bar corresponds to 2 μm. The microgels are labeled with core diameter along with their shell-to-core size ratio in parentheses.

**CONCLUSION**

In this work, we have investigated the isothermal compression of different colloidal monolayers assembled at air/water interfaces using a Langmuir trough in combination with a self-built setup for small-angle light scattering measurements (LT-SALS). This setup allowed us to measure the interparticle distances and characterize the structural order of the monolayers *in situ*, while the total available surface area was continuously reduced by the barriers of the Langmuir trough. When using core-shell microgels with rigid cores and soft and deformable shells, we found stark differences between microstructures analyzed *ex situ* (i.e., monolayers that were transferred to solid substrates) in comparison with the *in situ* structural analysis based on optical diffraction. The *ex situ* analysis revealed an "isostructural phase transition" from core-shell microgels in shell-shell contact to "core-core" contact during compression. In contrast, the *in situ* analysis revealed a continuous decrease of interparticle distance as the monolayer is compressed. No phase transition was observed. This key result was also confirmed by *in situ* real space analysis of the monolayer using fluorescence microscopy. As a proof of concept, we also demonstrated that the *in situ* investigation using small-angle light scattering can be also applied to monolayers of rigid particles as well as low optical contrast PNIPAM microgels.

LT-SALS is fast, non-destructive and relatively easy to set up from low cost components. Compared to *in situ* optical microscopy, it has several important advantages: 1) Very large monolayer areas > 1 mm$^2$ can be probed. Such large areas correspond to, for example, > $40 \times 10^4$ core-shell microgels that are simultaneously probed. 2) It is not necessary to have markers or strong refractive index contrast in colloid systems under investigation. 3) The measurement is less sensitive to external interferences such as vibrations. 4) Microstructural phase transitions become evident immediately due to changes of the diffraction pattern, i.e., a transition from a diffraction ring to Bragg peaks revealing the transition from a disordered to an ordered state. 5) The processing and analysis (e.g., radial averaging, peak position and width) of the diffraction patterns is much less prone to errors and less time-consuming when compared to real space analysis of microscopy images for which the centers of mass of all imaged particles have to be identified.

We believe that the presented methodology will stimulate further research on colloidal monolayers at liquid/liquid or air/liquid interfaces, in particular when softness and deformability of objects are studied.[50, 61] Furthermore, the fact that phase transitions can be directly monitored *in situ* at the respective interface will allow systematic studies required to achieve a more comprehensive understanding of colloidal assembly at interfaces[22] and enable on-demand tailoring of colloidal microstructures on solid substrates – provided that the transfer protocol to the substrate is suitable to maintain the microstructure. The next important steps in this line of research are further investigation on the role of the transfer protocol as well as the surface chemistry of the substrate during the drying procedure.

**MATERIALS**

Ethanol (Sigma-Aldrich, 99.8%), ethanol (Heinrich-Heine-University, central chemical storage, p.a.), tetraethyl orthosilicate (TEOS, Sigma-Aldrich, 98%), chloroform (Fischer Scientific, 99.8%), ammonium hydroxide solution ($NH_3$ (aq.), AppliChem, 32%), rhodamine B isothiocyanate (RITC, Sigma-Aldrich, mixed isomers), methacryloxyethyl thiocarbamoyl rhodamine B (MRB, Polysciences, Inc.), 3-aminopropyltrimethoxysilane (APS, Sigma-Aldrich, 97%), 3-(trimethoxysilyl)propyl methacrylate (MPS, Sigma-Aldrich, 98%), *N,N'*-methylenebisacrylamide (BIS, Sigma-Aldrich, 98%), potassium peroxodisulfate (KPS, Sigma-Aldrich, 99%), and sodium chloride (NaCl, Sigma-Aldrich, 99.5%) were used as received. Water was purified by Milli-Q system (18.2 MΩ cm$^{-1}$) and *N*-isopropylacrylamide (NIPAM, TCI, 97%) by recrystallization from cyclohexane (Fisher Scientific, 99.8%).

**SYNTHESIS**

**Silica particles and silica-PNIPAM CS microgels**

The detailed synthesis protocol for both silica nanoparticles and micron-sized silica-PNIPAM microgels can be found in.[59] In short, silica particles were synthesized via the well-known Stöber procedure. RITC dye was incorporated in the particles that were used for fluorescence microscopy experiments. The PNIPAM shell encapsulation was done via seeded precipitation polymerization.

The silica particles used to create monolayers at the air/water interface were measured to be 695 ± 22 nm (126 particles counted) in diameter by TEM. Its ethanolic dispersion was mixed with chloroform with 1:4 volume ratio to assist the spreading of the silica particles at the air/water interface. Surface charges were screened by adding 100 mM NaCl in the aqueous subphase of the Langmuir trough in order to achieve rigid sphere-like interactions.

The main CS microgels used for the *in situ* and *ex situ* comparison had a core with a diameter of 340 ± 20 nm. The $D_h$ of the total CS microgel was measured to be 920 ± 18 nm at 20 °C using dynamic light scattering (DLS). The purified dispersion was freeze-dried, re-dispersed in ethanol with 5 w/v% and stored on a 3D shaker overnight prior to the monolayer deposition at the air/water interface. Ethanol was used as spreading agent.

**PNIPAM microgel synthesis**

The synthesis protocol for the PNIPAM microgels was adopted from a previously published work.[62] 5 g of PNIPAM and 50 mg of BIS were dissolved in 50 ml of water in a three-neck round-bottom flask equipped with a reflux condenser and a magnetic stirrer. 1 mg of MRB dye was dissolved in 1 ml of water and added to the flask. The mixture was heated to 40 °C and purged with nitrogen while stirring. 20 ml of the mixture was transferred to another flask, where 10 ml of additional water was added. The rest was kept in a syringe with a needle and placed on a syringe pump for the continuous feeding of the monomers. The mixture in the flask was heated to 80 °C and equilibrated. 10.4 mg of KPS was dissolved in 2 ml water and added to the flask. Once the dispersion started to become turbid, indicating that the polymerization was initiated, the syringe pump was started with the speed of 1 ml/min. 5 minutes after the feeding process, the polymerization was quenched by dipping the flask in an ice bath and the dispersion was filtered through glass wool. The synthesized PNIPAM microgels were dialyzed against water for two weeks, freeze-dried and re-dispersed in ethanol (1 w/v%) as for the CS microgels. The size of the PNIPAM microgels at 20 °C was determined by DLS ($D_h$ = 858 nm ± 41 nm).

**EXPERIMENTAL**

***Ex situ* investigation after Langmuir-Blodgett deposition (Method 1)**

For the *ex situ* analysis, we followed the well-established protocol to study the phase behavior of CS microgel monolayers at the air/water interface.[12, 29, 30] According to the protocol, the microgel monolayer at the air/water interface is simultaneously transferred to a substrate during the compression, dried as the substrate is pulled out and examined under a microscope, hence referred to as an *ex situ* approach. The total duration of the substrate pulled out is often matched with the total duration of the compression, consequently enabling the position of the substrate to the corresponding Π tracing. This link between the substrate position to Π was established under the assumption that the number of particles transferred from the air/water interface per time is negligible thus does not influence the measured value of Π.

The transfer of the monolayer was carried out using a Langmuir-Blodgett deposition trough (Microtrough G2, Kibron Inc.) equipped with a film balance, two Delrin barriers, a dip coater and an acrylic cover box. A standard microscope glass slide was treated in an ultrasonic bath sequentially with Hellmanex aqueous solution (2 vol%), water (2 ×) and in ethanol (2 ×) for 15 minutes each. The cleaned glass slide was then cut in half along its length and the position markings were carved on its back to trace the corresponding Π at the moment of monolayer transfer (see **Figure 10A** for a schematic illustration of the procedure). Before the deposition of the particle monolayer, the trough and the barriers were thoroughly cleaned with water, ethanol and again rinsed with water. The trough was then filled with water with the barriers closed. The glass slide was thoroughly rinsed with water before being mounted to the dip coater (parallel to the barriers and perpendicular to the air/water interface), positioned at the center of the trough and lowered 55 mm below the interface. An aspirator with a narrow tip was used to remove any residual floating substances at the interface between the two barriers as well as to flatten the interface by lowering its level to the height of the trough wall. A Wilhelmy plate was rinsed with water and ethanol and held over a flame to remove any impurities and cooled before it was mounted to the film balance. The barriers were then opened to the maximum area. Only when the fluctuation of the surface tension value was below 0.3 mN/m while opening the barriers, the particles were deposited at the air/water interface using a 10 or 20 μl micropipette. The colloidal dispersion was treated alternating between vortex mixing and sonication for 2-4 minutes prior to the deposition. The injection was done slowly at a shallow angle with the tip of the micropipette gently touching the interface. The compression was started with the speed of 150 mm$^2$/min after at least 15 minutes of equilibration time after deposition. The glass substrate was pulled out simultaneously at the speed of 84 mm/min over a time that matched the total duration of the compression. The glass substrate was left hanging until it was completely dried. Three images were taken every 2.5 mm using an upright microscope (Eclipse LV150N, Nikon). The acquired images were processed and analyzed by ImageJ (1.53k, National Institutes of Health, USA) alongside the recorded compression isotherm. Additional information regarding image processing can be found in **SI**. All experiments were done at room temperature. The interparticle distances were determined from the first peak of the radial distribution functions (RDFs) and are denoted as $D_{c-c}^{im,ex}$, where $D$ stands for "distance", *im* for "image", *ex* for "*ex situ*" and *c-c* for "center-to-center".

***In situ* investigation by fluorescence microscopy combined with a Langmuir trough (Method 2)**

The *in situ* measurements by fluorescence microscopy were conducted in another Langmuir trough (KSV NIMA inverted, Biolin Scientific) equipped with two Delrin barriers, an inverted microscopy trough and a black acrylic cabinet. A microscope (Olympus IX73) equipped with a mercury lamp, a fluorescence filter set, a CMOS camera and a 60× objective was used to probe the colloid-laden interface at various values of Π. The setup was placed on an optical table combined with pneumatic vibration isolation (Nexus, Thorlabs Inc.) and is illustrated in **Figure 10B**. The microscopy trough was cleaned and prepared in the same way as for the *ex situ* experiments previously described. Once the air/water interface was clean enough (surface pressure fluctuation below 0.3 mN/m, measured by a Wilhelmy plate), the microgels were deposited at the interface. The compression was done stepwise with the compression speed of 10 mm/min. Images were taken after at least 15 minutes of equilibration time at each step. Three different volumes of the microgel dispersion were used and consequently three compression isotherms were measured to address the full range of Π. Three images were acquired from different positions for each measured Π and processed by ImageJ alongside the Π measured at the moment of image acquisition. All experiments were done at room temperature.

*In situ* **investigation by LT-SALS (Method 3)**

The *in situ* measurements by LT-SALS were performed with the same Langmuir trough (KSV NIMA) used for the *in situ* fluorescence microscopy experiments. Along with the Langmuir trough, the cabinet was placed on an optical table, on which a blue diode laser (MediaLas, LDM-20-405, 20 mW, 405 nm) and two mirrors (Thorlabs, BB1-E02) were set up. A customized metal frames was installed around the Langmuir trough to mount a paper screen and a CCD camera (Thorlabs, DCU223C-MVL6WA) above the trough. The schematic of the setup is depicted in **Figure 10C** (see also the **SI** for more details regarding the setup and laser alignment). The monolayer deposition at the air/water interface was done as for the *in situ* fluorescence microscopy study. The compression was started after at least 15 minutes of equilibration time with the speed of 150 mm$^2$/min. The laser intensity was adjusted with a neutral density filter (Thorlabs, NDC-50C-4M). Diffraction patterns were recorded during the entire compression at 0.2 frames per second. The resulting video was analyzed alongside the recorded compression isotherm using ImageJ (one data point every five seconds). All experiments were done at room temperature.

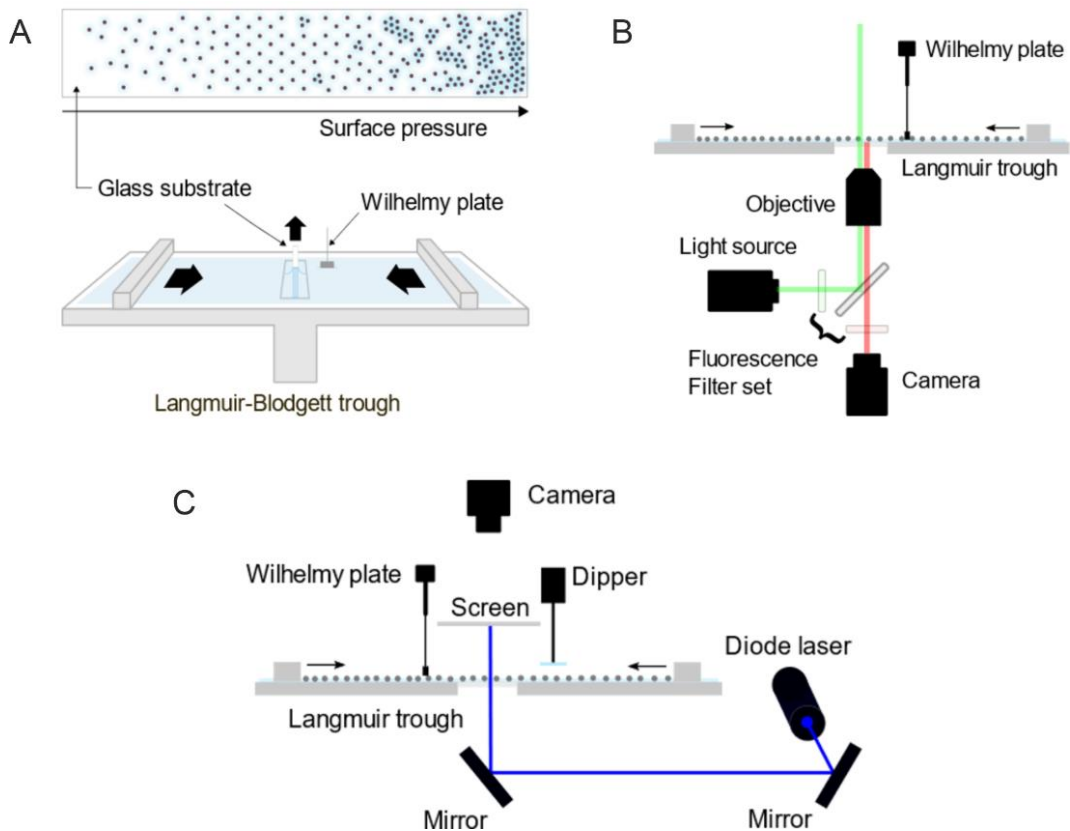

**Figure 10.** Schematic illustration of A) the transfer of a colloidal monolayer to a glass substrate using a Langmuir-Blodgett trough, B) the Langmuir trough combined with a fluorescence microscope, and C) the Langmuir trough combined with a small-angle light scattering (LT-SALS) setup.

**Dynamic light scattering (DLS)**

DLS measurements were performed with a 3D LS spectrometer (LS Instruments) at a constant temperature of 20 °C. The measurements were repeated three times with 40 seconds of acquisition time. The device was equipped with a HeNe laser (632.8 nm), a decalin bath and two avalanche photodiodes in pseudo-cross-correlation mode as detectors. The dilute samples (volume fraction $\ll 0.001$) were measured in borosilicate cuvettes with an outside diameter of 10 mm. The obtained intensity-time autocorrelation functions were analyzed using cumulant analysis.

**Transmission electron microscopy (TEM)**

TEM measurements were performed using a JEOL JEM-2100 Plus microscope operated in bright-field mode at 80 kV acceleration voltage. The sample preparation was done by applying a drop of the respective particle dispersion on a carbon-coated copper grid (200 mesh, Science Services) and drying at room temperature. The captured images were then processed with ImageJ for the size analysis.

**SUPPORTING INFORMATION**

Additional comparison of *ex* and *in situ* measurement, interparticle distance calculation from diffraction images, LT-SALS measurement with different wavelength lasers, LT-SALS setup and monolayer preparation for LT-SALS, image processing steps and other details (PDF), a video of LT-SALS measurement (AVI).

**AUTHOR INFORMATION**

**Corresponding Author**

Name: Matthias Karg*

E-mail: karg@hhu.de

**NOTES**

The authors declare no competing financial interest.

**ACKNOWLEDGEMENTS**

The authors acknowledge the DFG and the state of NRW for funding the cryo-TEM (INST 208/749-1 FUGG). M.K. acknowledges the German Research Foundation (DFG) for funding under grant KA3880/6-1. The authors also would like to thank Sonja Schiller (HHU Düsseldorf) for assistance in building the setup, Dr. Kiran Kaithakkal Jathavedan for providing the PNIPAM microgels, Marius Otten (HHU Düsseldorf) for TEM imaging and Jonathan Garthe (HHU Düsseldorf) for Langmuir-Blodgett deposition of many samples. K.K would like to express special thanks to Prof. Michel Cloitre, Prof. Igor Potemkin, Prof. Jérôme J. Crassous, Prof. Janne-Mieke Meijer, Prof. Walter Richtering and others for inspiring lectures and fruitful discussions during the 3$^{rd}$ International Summer School on functional microgels and microgel systems. The authors acknowledge the reviewers for their constructive feedback and helpful suggestions.

# Compression of colloidal monolayers at liquid interfaces: *in situ* vs. *ex situ* investigation

*Keumkyung Kuk,[a] Vahan Abgarjan,[a] Lukas Gregel,[a] Yichu Zhou,[a] Virginia Carrasco Fadanelli,[b] Ivo Buttinoni,[b] and Matthias Karg [a*]*

[a]Institut für Physikalische Chemie I: Kolloide und Nanooptik, Heinrich-Heine-Universität Düsseldorf, Universitätsstr. 1, 40225 Düsseldorf, Germany

[b]Institut für Experimentelle Physik der kondensierten Materie, Heinrich-Heine-Universität Düsseldorf, Universitätsstr. 1, 40225 Düsseldorf, Germany

***In situi* fluorescence microscopy of the main CS microgel at low Π**

**Figure S1** shows a microscopy image (*in situ*) of a monolayer of the main CS microgel (the same CS microgel presented in **Figure 1-4** – the main CS microgel). Here, Π is very low (0.2 mM/m) but still showing that attractive interactions are present at the interface.

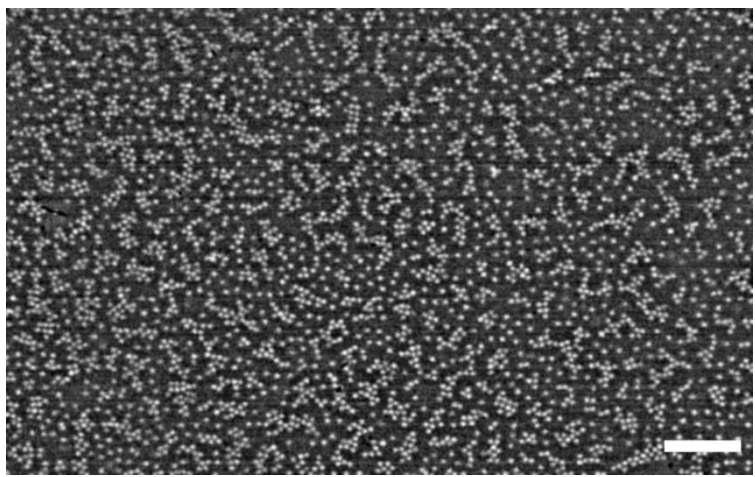

**Figure S1.** Fluorescence microscopy image (**Method2**) of a CS microgel monolayer at Π = 0.2 mM/m. The scale bar corresponds to 20 μm. The image was processed by ImageJ (Bandpass filter – Gaussian Blur) for better visibility.

**Additional *ex situ* vs *in situ* differences for the main CS microgels**

**Figure S2A** shows $D_{c-c}$ plotted as a function of the particle number per unit area ($n_P/A$). Theoretical values of $D_{c-c}$ for close-packed, perfectly hexagonally ordered monolayers were calculated according to our already published work.[59] These values are compared with *ex* and *in situ* microscopy data. **Figure S2B** illustrates the systematic shift of $n_P/A$ in *ex situ* measurement compared (filled squares) to the *in situ* method (open squares) at the same measured Π.

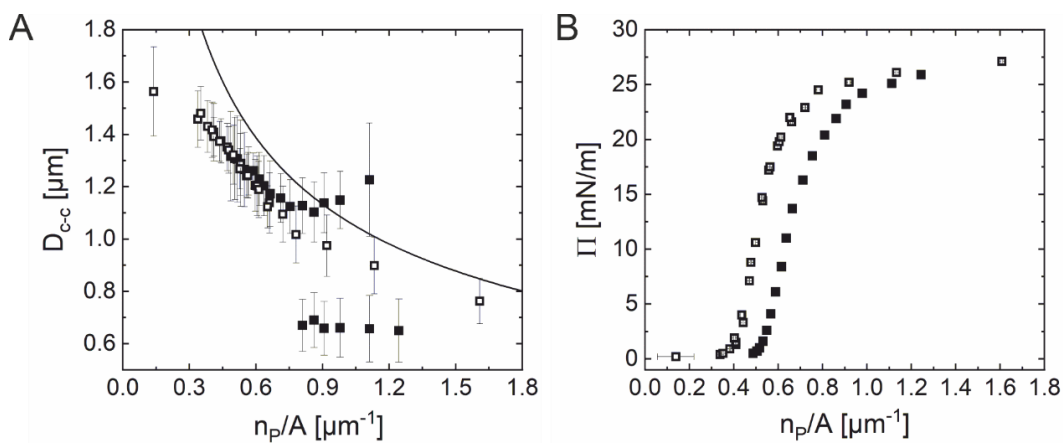

**Figure S2**. A) $D_{c-c}$ as a function of $n_P/A$. The solid line corresponds to the calculated, theoretical evolution of $D_{c-c}$ (under the assumptions that the monolayer has a perfect hexagonal symmetry with the area fraction of 0.9069). The filled squares correspond to the *ex situ* and the empty squares to the *in situ* measurement results, respectively. B) $\Pi$ as a function of $n_P/A$ obtained from *ex situ* (filled squares) and *in situ* (empty squares) measurements.

**Monolayers of silica particles**

The monolayer of silica particles shown in **Figure 7A** of the main manuscript was transferred and dried onto a solid substrate using Langmuir-Blodgett deposition. **Figure S3** shows microscopic images and SALS patterns of these monolayers for different values of $\Pi$ (see caption). The diffraction patterns vary from ring-like to peak-like depending on the position on the substrate.

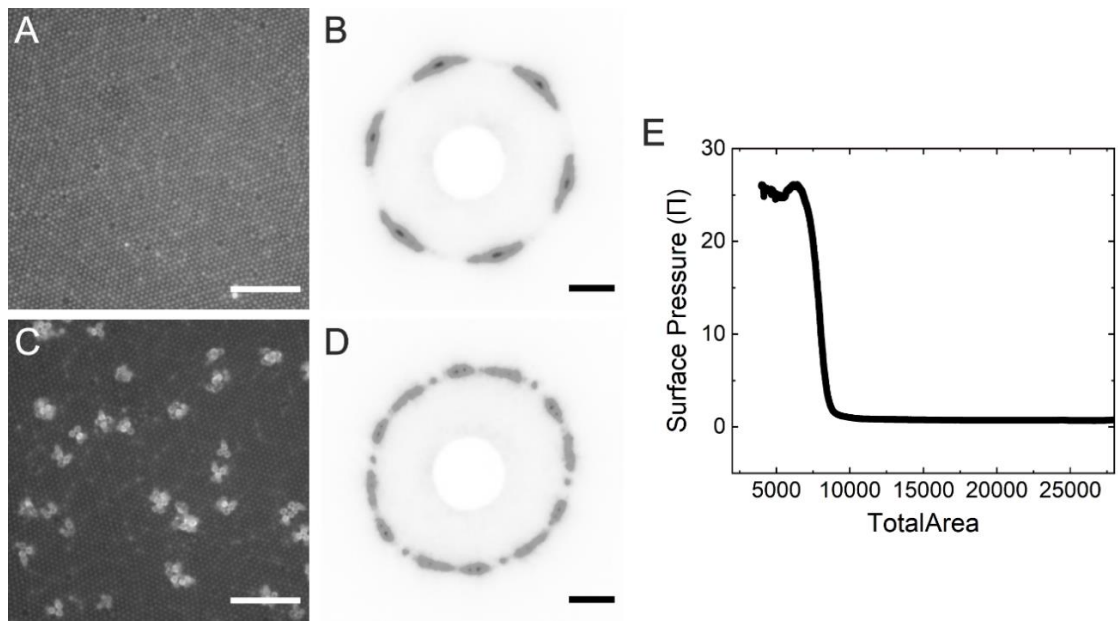

**Figure S3**. A) Optical light microscopy image and B) diffraction pattern of a dried silica monolayer (*ex situ*) at $\Pi = 0.5$ mN/m. C) and D) show the same but at $\Pi$ approx. 20 mN/m. The scale bars in A) and C) correspond to 10 μm. The scale bars in B) and D) correspond to 10 mm. E) Compression isotherm recorded during the Langmuir-Blodgett deposition.

**Monolayers of (coreless) PNIPAM microgels**

The system of coreless microgels shown in **Figure 7B** of the main manuscript was also investigated *in situ* using fluorescence microscopy (**Method 2**), also in a range of Πs where the interparticle distance is too large to be resolved in our current LT-SALS setup. **Figure S4** shows microscopic images of the microgel monolayer at different Π. We do not observe an "isostructural phase transition".

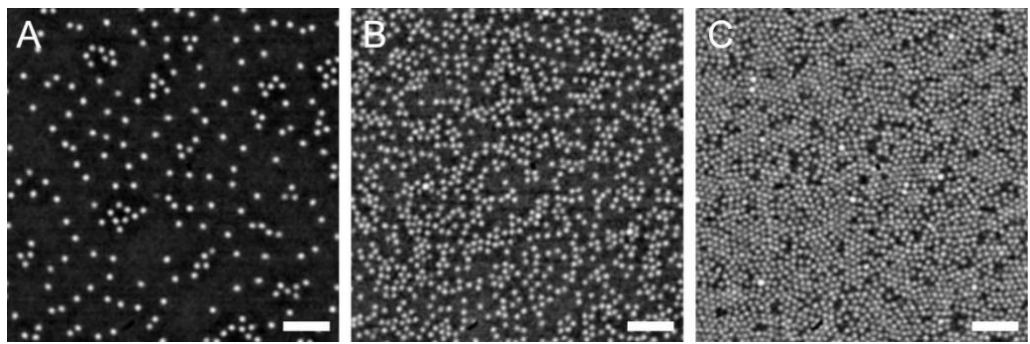

**Figure S4**. Microgel monolayer (*in situ*) Π at A) 3.7, B) 28.4, C) 29.2 mN/m. The scale bars correspond to 10 μm. The images were processed by ImageJ (Bandpass filter – Gaussian Blur) for better visibility.

The same monolayer of microgels was then studied *ex situ* after Langmuir-Blodgett deposition. **Figure S5A** and **C** show optical light microscopy images of the dried microgel monolayers prepared at two different Π of 30.2 (black) and 31.2 mN/m (red). **Figure S5B** and **D** show AFM images of the corresponding microgel monolayers. **Figure S5E** shows calculated radial distribution functions (RDFs) as well as nearest neighbor center-to-center distances, $D_{c-c}$. **Figure S5F** displays the diffraction pattern recorded by SALS, representative for the microgel monolayers taken from Π between 30.2 and 31.2 mN/m.

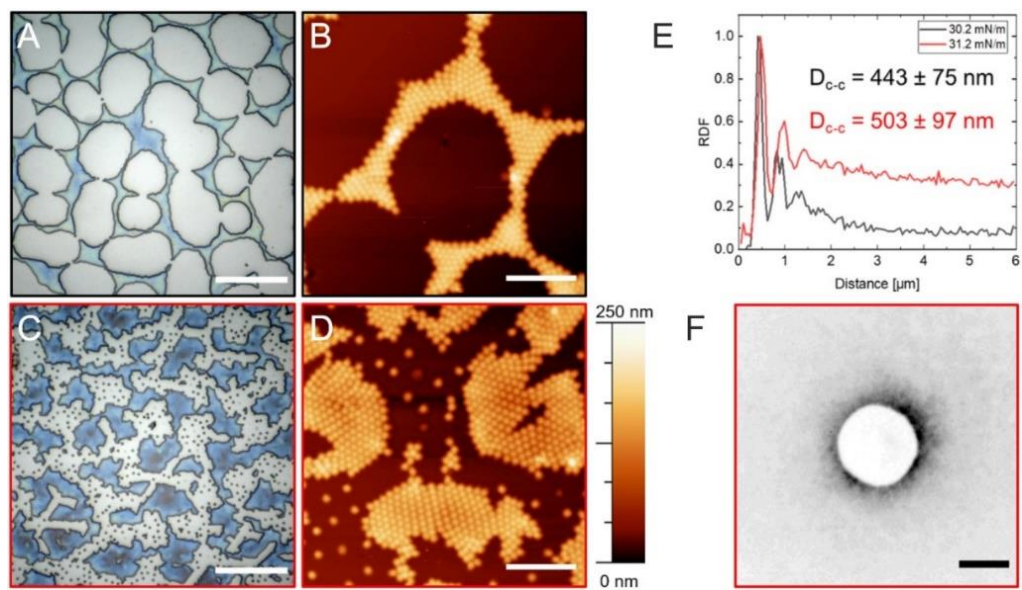

**Figure S5**. A) Microgel monolayer (*ex situ*) at Π = 30.2 mN/m by light microscopy, scale bar: 20 μm B) by AFM, scale bar: 5 μm. C)-D) the same data set for the microgel monolayer (*ex situ*) at Π = 31.2 mN/m. E) interparticle distance from RDF. F) Scattering pattern of the microgel monolayer from C), scale bar: 10 mm.

**AFM images of CS microgel from Figure 9 –** 105 nm (4.8)

The microscopic image of CS microgel ($D_c$: 105 nm, shell-to-core size ratio: 4.8) in **Figure 9C2** (dried in open air) of the main manuscript could not be resolved due to the relatively homogeneous refractive index of the microgel, in comparison with **Figure 9C1** (dried with heat gun). **Figure S6A** and **C** show AFM measurements on these monolayers at lower Π (10 mN/m), and **Figure S6C** and **D** are the corresponding phase images.

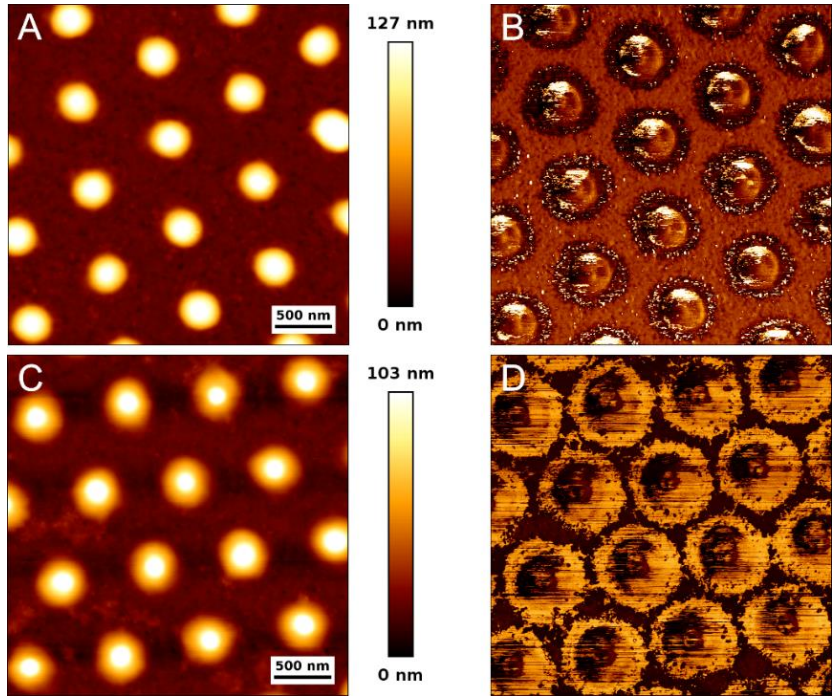

**Figure S6**. A) AFM image of CS microgel – 105 nm (4.8) – monolayer dried with heat gun B) phase image of A). C)-D) the same set of data for the monolayer dried in open air at room temperature.

**Interparticle distance from LT-SALS**

The interparticle distance measured by LT-SALS ($D_{c-c}^{SALS}$) was calculated as below:

$$q = \frac{4\pi n}{\lambda} \sin\left(\frac{1}{2} atan\left(\frac{xy}{D_{S-D}}\right)\right)$$

where $q$ is the magnitude of the scattering vector in µm$^{-1}$, $n$ the refractive index (refractive index of air, $n = 1$), $\lambda$ the wavelength of the light in µm, $x$ the distance from primary beam in pixel, $y$ the conversion factor in mm per pixel and $D_{S-D}$ the sample-to-detector distance in mm. The scattering vector yields the lattice spacing $D_{hk} = \frac{2\pi}{q}$. For a two-dimensional, hexagonally ordered system, the interparticle distance is $D_{c-c} = \frac{2}{\sqrt{3}} d_{hk}$. With a blue diode laser ($\lambda$ = 405 nm), the available $q$- value ranges from 0.39 to 15.74 µm$^{-1}$ ($D_{c-c}$ approximately from 460 nm up to 18 µm). With a green and red diode laser ($\lambda$ = 532 nm, 632.8 nm), the available $q$-value ranges from 0.30 to 11.98 µm$^{-1}$ ($D_{c-c}$ approximately from 600 nm up to 24 µm) and from 0.25 to 10.07 µm$^{-1}$ ($D_{c-c}$ approximately from 720 nm up to 29 µm), respectively. **Figure S7** depicts LT-SALS measurements done with blue and red lasers as well as the $D_{c-c}$ range for three different laser types graphically.

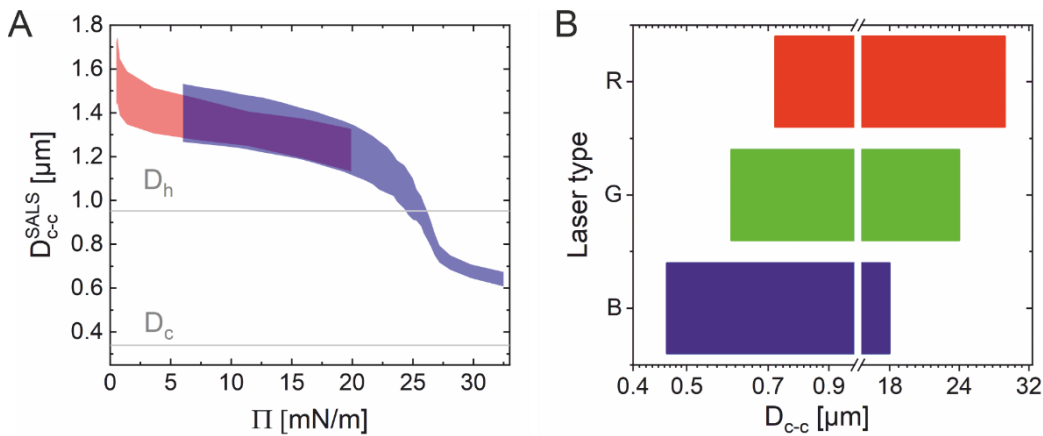

**Figure S7**. A) LT-SALS measurement done with blue and red lasers on the same CS microgel. B) Calculated possible range of $D_{c-c}$ for different laser wavelengths for our setup, R-red, G-green and B-blue.

**Image processing and analysis by ImageJ (1.53k, National Institutes of Health, USA)**

Radial Distribution Functions (RDFs) were calculated with ImageJ macro version 2011-08-22 by Ajay Gopal using the center of mass positions of each CS microgel. To find the centers, the *ex situ* light microscopy images were pre-processed with Gaussian blur. Particles at the edges of the images were excluded. The *in situ* fluorescence microscopy images were then processed with Bandpass filter, background subtraction and Gaussian blur. Grey scaled LT-SALS images were radially and azimuthally averaged by Radial Profile Plot (Version 2009-08-14 by Paul Baggethun) and Azimuthal Average (Version 2007-09-08 by Philippe Carl), respectively.

**LT-SALS setup**

The level of accuracy was checked with all the involved components in the laser path using a circular level. The laser was aligned with the camera center with two mirrors and through the microscopy window of the trough by using a pinhole on a rail, which consisted of two parallel rods screwed into the optical plate. After the alignment, the rail and the pinhole were removed. The Langmuir trough and the camera were placed back in the laser path. The paper screen (width: 90 mm) was rolled around two metal rods, fastened parallel to the trough and fixed on the customized frame, see **Figure S8**. The laser beam center was marked on the screen for various size of beam stops to be glued on when required. The sample-to-detector distance ($D_{S\text{-}D}$) was measured with a ruler ensuring all four corners of the screen have the same distance to the trough wall. The pixel/mm value was determined using millimeter paper after all the involved components were fixed on their positions. The screen was rolled back and put aside on one metal rod for the cleaning of the trough. The trough was filled again with water before the screen was rolled out and fixed back to its position. Then the monolayer was deposited at the air/water interface. The $D_{S\text{-}D}$ of our current setup could be varied from 25 to 200 mm (scattering angle ranges from 2 - 74°).

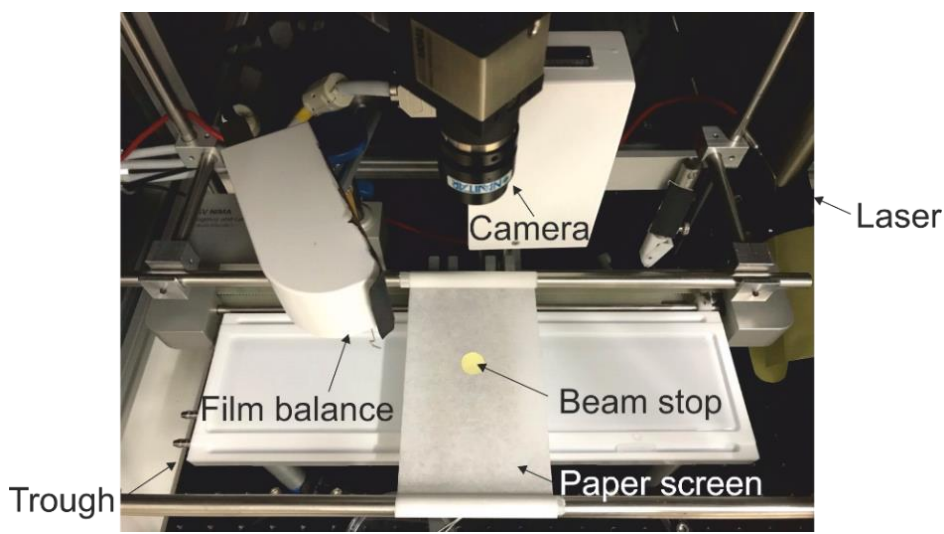

**Figure S8**. Photograph of the LT-SALS setup.